\documentclass[12pt]{article}
\usepackage{graphics}

\newlength{\dinwidth}
\newlength{\dinmargin}
\def\lapproxeq{\lower .7ex\hbox{$\;\stackrel{\textstyle<}{\sim}\;$}}
\def\gapproxeq{\lower .7ex\hbox{$\;\stackrel{\textstyle>}{\sim}\;$}}
\def\be{\begin{equation}}
\def\ee{\end{equation}}
\def\bea{\begin{eqnarray}}
\def\eea{\end{eqnarray}}
\newcounter{bean}
\begin{document}
\begin{flushright}
IPPP/01/05 \\
DCPT/01/10 \\
31 January 2001
\end{flushright}

\vspace*{2cm}             

\begin{center}                                                         
{\Large \bf Unintegrated parton distributions}

\vspace*{1cm}
M.A. Kimber$^a$, A.D. Martin$^a$ and M.G. Ryskin$^{a,b}$

\vspace*{0.5cm}
$^a$ Department of Physics, University of Durham, Durham, DH1 3LE \\
$^b$ Petersburg Nuclear Physics Institute, Gatchina, St.~Petersburg,
188300, Russia
\end{center}

\vspace*{1cm}

\begin{abstract}
We describe how to calculate the parton distributions
$f_a (x, k_t^2, \mu^2)$, unintegrated over the parton transverse momentum
$k_t$, from auxiliary functions $h_a (x, k_t^2)$, which satisfy
single-scale evolution equations.  The formalism embodies both DGLAP and
BFKL contributions, and accounts for the angular ordering which comes
from coherence effects in gluon emission.  We check that the unintegrated
distributions give the measured values of the deep inelastic structure
function $F_2 (x, Q^2)$.
\end{abstract}

\section{Introduction}   
   
Conventionally deep inelastic lepton-proton scattering is described in terms
of scale-dependent parton distributions, $a (x, \mu^2)$, where $a = xg$ or
$xq$.  These distributions correspond to the density of partons in the proton
with longitudinal momentum fraction $x$, integrated over transverse momentum
up to $k_t = \mu$.  They satisfy DGLAP evolution in $\mu^2$.  The kinematic
region $k_t < \mu$ gives the leading $\ln \mu^2$ approximation to deep
inelastic scattering.

For less inclusive processes it is, however, necessary to consider
distributions unintegrated over the transverse momentum $k_t$ of the parton.
The unintegrated distributions have the advantage that they exactly correspond
to the quantity which enters the Feynman diagrams and therefore allow for the
true kinematics of the process even at leading order (LO).  These
distributions $f_a (x, k_t^2, \mu^2)$ depend on two hard scales\footnote{This
property is hidden in the conventional distributions as $k_t$ is integrated up
to the scale $\mu$.}:  $k_t$ and the scale $\mu$ of the probe.  The scale
$\mu$ plays a dual role.  On the one hand it acts as the factorization scale,
while on the other hand it controls the angular ordering of the partons
emitted in the evolution \cite{CCFM}.

Clearly it is desirable to also include $\ln (1/x)$ BFKL-type contributions
in the evolution.  Recall that both DGLAP and BFKL evolution are essentially
equivalent to ordered evolution in the angles of the emitted
partons\footnote{At LO we have strong ordering of the emission angles, $\ldots
\theta_i \ll \theta_{i + 1} \ldots$; on the other hand if, at one step of the
evolution $\theta_i \sim \theta_{i + 1}$, then this contribution is included
inside the NLO splitting function.}.  In the DGLAP collinear approximation the
angle increases due to the growth of $k_t$, while in BFKL the angle $(\theta
\simeq k_t/k_\ell)$ grows due to the decreasing longitudinal momentum fraction
as we proceed along the emission chain from the proton.  The factorization
scale $\mu$ separates the partons associated with emission from different
parts of the process, that is from the beam and target protons (in $pp$
collisions) and from the hard subprocess.  For example it separates emissions
from the beam (with polar angle $\theta \lapproxeq 90^\circ$) from those from
the target (with $\theta \gapproxeq 90^\circ$), and from the intermediate
partons from the hard subprocess.  This separation was proved in
Ref.~\cite{CCFM} and originates from the destructive interference of the
different emission amplitudes in the angular boundary regions.  If the
longitudinal momentum fraction is fixed by the hard subprocess, then the
limits on the angles can be expressed in terms of a factorization scale $\mu$
which corresponds to the upper limit\footnote{The $t$-channel parton may have
$k_t$ up to $\mu/z$, characteristic of BFKL effects, whereas for LO 
DGLAP the $s$ and $t$-channel partons are both limited by $k_t < \mu$.
Of course, some $k_t>\mu$ contribution will arise from the NLO splitting
functions.}
on the allowed values of the ($s$-channel) parton $k_t$.

Since the parton distributions depend on two scales we potentially have to
deal with complicated (CCFM \cite{CCFM}) evolution 
equations for the $f_a (x, k_t^2, \mu^2)$
functions.  Of course it is possible to work with two-scale distributions,
but this is much more complicated \cite{TS} and up to now has only 
proved practical with
Monte Carlo generators \cite{JUNG}.
However, the evolution process is essentially controlled by one
quantity, the emission angle, and on this basis we may expect to be able to
obtain the distributions $f_a (x, k_t^2, \mu^2)$ from single-scale evolution
equations.  Therefore it should be possible to follow an analytic approach
where the physical assumptions are much more evident and where, in principle,
NLO corrections can be included.  Moreover, in practice, it is much easier
to use the same unintegrated distributions 
to describe different hard processes and
to perform global parton analyses.

The outline of this paper is as follows.  The key observation is that the
$\mu$ dependence of the unintegrated distributions enters at the last step
of the evolution, and so we may use single-scale evolution equations.
The procedure is first described in Section~2 in the case of pure DGLAP
evolution, and 
then extended to include $\ln (1/x)$ effects in Section~3.  In the latter
case we use the solution of a single-scale equation which unifies DGLAP
and BFKL evolution \cite{KMS}, and perform a final evolution step which brings
in the dependence on the second scale.  Ref.~\cite{KKMS} also
generated the two-scale unintegrated gluon from the same unified evolution
equation, but with a different procedure\footnote{In this work we impose the
angular ordering constraint in both the BFKL and DGLAP terms, and as
a result do not have
an exact equality between the integral up to $\mu^2$ of the
unintegrated distributions and the value of the integrated distribution.
Ref.~\cite{KKMS} takes the opposite approach; that is, exact equality
with the integrated distribution
is imposed and as a result angular ordering of the BFKL contribution
is not complete. The difference is a NLO effect.}.  The unintegrated 
gluons obtained
using the procedures described in Sections~2, 3 and Ref. \cite{KKMS} are
compared in Section~4.  In Section~5 we describe how the structure
function $F_2$ is calculated from the unintegrated parton distributions,
and in Section~6 we discuss the relationship between the unintegrated
and integrated distributions.  Finally in Section~7 we give our conclusions.  

\section{Unintegrated DGLAP partons}   
   
It is informative to review how unintegrated distributions $f_a (x, k_t^2,
\mu^2)$ may be calculated from the conventional (integrated) parton densities,
$a (x, \mu^2)$, in the case of pure DGLAP evolution.  The procedure was
explained in Ref.~\cite{KMR}.  We start from the DGLAP equation\footnote{For
the $g \rightarrow gg$ splitting we have to insert a factor $z^\prime$ in
front of $P_{gg} (z^\prime)$ in the last integral of (\ref{eq:a1}) to account
for the identity of the produced gluons.}
\be   
\label{eq:a1}   
\frac{\partial a (x, \mu^2)}{\partial \ln \mu^2} \; = \; \frac{\alpha_S}{2
\pi} \:  \left [ \int_x^{1 - \Delta} \:  P_{aa^\prime} (z) \:  a^\prime \left
( \frac{x}{z}, \mu^2 \right ) dz \:  - \:  a (x, \mu^2) \:  \sum_{a^\prime} \:
\int_0^{1 - \Delta} \:  P_{a^\prime a} (z^\prime) dz^\prime \right ]
\ee   
where in the first term a sum over all possible parent partons $a'$ is
implied.  This first term on the right-hand-side describes the number of
partons $\delta a$ emitted in the interval $\mu^2 < k_t^2 < \mu^2 + \delta
\mu^2$.  Such emission clearly changes the transverse momentum $k_t$ of the
evolving parton.  If we were to neglect the virtual contribution in
(\ref{eq:a1}), then the unintegrated parton density would be given simply by
\bea   
\label{eq:a2}   
f_a (x, k_t^2) & = & \left .  \frac{\partial a (x, \mu^2)}{\partial \ln \mu^2}
\right |_{\mu^2 = k_t^2} \nonumber \\ & & \nonumber \\ & = & \frac{\alpha_S
(k_t^2)}{2 \pi} \:  \int_x^{1 - \Delta} \:  P_{aa^\prime} (z) \:  a^\prime
\left ( \frac{x}{z}, k_t^2 \right ) dz.
\eea   
The virtual contribution in (\ref{eq:a1}) does not change the parton $k_t$ and
may be resummed to give the survival probability $T_a$ that parton $a$ with
transverse momentum $k_t$ remains untouched in the evolution up to the
factorization scale.  The survival probability is given by
\be   
\label{eq:a3}   
T_a (k_t, \mu) \; = \; \exp \left (-\int_{k_t^2}^{\mu^2} \:  \frac{\alpha_S
(k_t^{\prime 2})}{2 \pi} \:  \frac{dk_t^{\prime 2}}{k_t^{\prime 2}} \:
\sum_{a^\prime} \:  \int_0^{1 - \Delta} \:  P_{a^\prime a} (z^\prime) \:
dz^\prime \right ),
\ee  
 \`{a} la Sudakov form factor.  Thus the probability to find parton $a$ with
transverse momentum $k_t$ (which initiates a hard subprocess with
factorization scale $\mu$) is
\be   
\label{eq:a4}
f_a (x, k_t^2, \mu^2) \; = \; T_a (k_t, \mu) \left [ \frac{\alpha_S (k_t^2)}{2
\pi} \:  \int_x^{1 - \Delta} \:  P_{aa^\prime} (z) \:  a^\prime \left (
\frac{x}{z}, k_t^2 \right ) dz \right ].
\ee   
It is at this {\it last step of the evolution} that the unintegrated
distribution becomes dependent on the two scales, $k_t^2$ and $\mu^2$.

We now have to take care to specify the value of the infrared cut-off
$\Delta$, which is introduced to protect the $1/(1 - z)$ singularity in the
splitting functions arising from soft gluon emission.  In the original DGLAP
equation, (\ref{eq:a1}), which describes the evolution of the integrated
distributions, this singularity is cancelled between the real emission and
virtual contributions.  However after the resummation of the virtual terms,
the real soft gluon emission must be accounted for explicitly since it changes
the $k_t$ of the parton.  Thus we have to find the physically appropriate
choice of the cut-off $\Delta$ to provide the angular ordering of the gluon
emissions\footnote{Although the splitting functions $P_{gq}$ and $P_{qg}$ are
not singular at $z = 1$ it is natural to use the same prescription for both
the quark and the gluon distributions.}.

In Ref.~\cite{KMR} the cut-off was taken to be $\Delta=k_t/\mu$.  As a
consequence the two-scale unintegrated distributions $f_a(x,k_t^2,\mu^2)$
of \cite{KMR} vanish for $k_t>\mu$,
in accordance with the DGLAP strong ordering in $k_t$.
However we can do better and impose the more correct angular ordering in 
the last step of the evolution. It was shown in Refs.~\cite{CCFM,KKMS} that
this leads to a constraint on the scale $\mu$, namely
\be   
\label{eq:a12}   
\Theta (\theta - \theta^\prime) \;\: \Rightarrow 
\;\: \mu \; > \; z k_t/(1 - z).
\ee
Thus the maximum allowed value of the integration variable $z$ is
\be
\label{eq:a5} z_{max} \; = \; \frac{\mu}{\mu+k_t}
\ee
and the corresponding cut-off $\Delta= k_t/(\mu+k_t)$.  
Of course the same $\Delta$ must be used both in
the real emission integral in (\ref{eq:a4}) and in the survival
probability $T$ in (\ref{eq:a3}).\footnote{In equation (\ref{eq:a3}),
$\Delta=k'_t/(\mu+k'_t)$ is the appropriate cut-off for $z'$.} In fact
we shall see that the imposition of angular ordering at the last step
of the evolution leads to physically reasonable parton $k_t$ distributions
which extend smoothly into the domain $k_t>\mu$.

\begin{figure}\centering
\scalebox{0.5}{\resizebox{\textwidth}{!}{\includegraphics{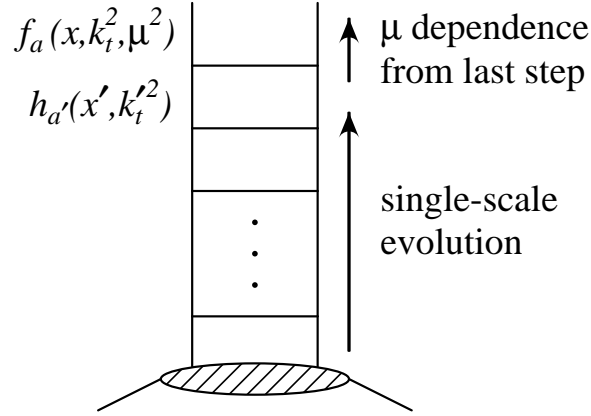}}}
\caption{An illustration of our procedure, in which the evolution of a
single-scale unintegrated parton is followed by a final step of the ladder
which introduces dependence on the second hard scale, $\mu$.}
\label{fig:laststep}
\end{figure}

\section{Inclusion of \boldmath{$\ln (1/x)$} effects}

We wish to generalize the above method to include the leading $\ln (1/x)$
contributions.  Clearly there can be different forms of the `unified'
evolution equation summing up the leading DGLAP and BFKL logarithms, where
the ambiguity is at the subleading level.  The aim is to find a good
prescription which is not too complicated, but which can account for all the
physically relevant kinematic effects just at LO level.  In other words we
seek an equation which sums up the major part of the subleading corrections in
a LO framework.

Let us consider, for the moment, just the gluon distribution.  Recall that the
unintegrated distribution $f (x, k_t^2, \mu^2)$ depends on two scales.  As in
the pure DGLAP case of Section~2 we wish to work in terms of a single-scale
evolution equation, and then to restore the scale $\mu$, and the full
kinematics, at the last step of the evolution.  This is illustrated
schematically in Figure~\ref{fig:laststep}.  For an analysis which
incorporates BFKL effects, the appropriate single-scale distribution is the
auxiliary function
\be
\label{eq:a6}
h (x, k_t^2) \; = \; \frac{\partial (xg (x, k_t^2))}{\partial \ln k_t^2}.
\ee
Note that $h (x, k_t^2)$ is precisely the function which satisfies the BFKL
equation in the low $x$ limit.

\begin{figure}\centering
\scalebox{0.7}{\resizebox{\textwidth}{!}{\includegraphics{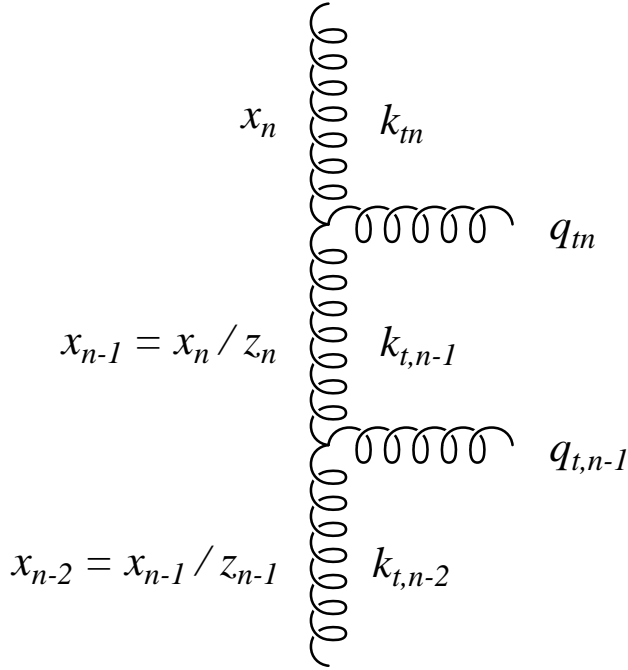}}}
\caption{part of the evolution chain.  We commonly write $k_t$ for $k_{tn}$
and then the parent's transverse momentum as $k'_t$.  The radiated transverse
momentum is $q_t$.  Unified evolution is naturally performed \cite{KKMS}
in terms of the rescaled transverse momentum $q_n=q_{tn}/(1-z_n)$.}
\label{fig:vars}
\end{figure}   
   
Both BFKL and DGLAP evolution correspond to angular ordering of the emission
angles, and are single-scale equations.  At LO, strong angular ordering
automatically comes either from strong ordering in $x$ ($z \ll 1$) for BFKL or
from strong $k_t$ ordering $(k_t^{\prime 2} \ll k_t^2)$ for DGLAP.  For the
unintegrated gluon, we face a problem when $k_t \sim \mu$ in the DGLAP
framework, and similarly we have a problem when $z \sim 1$ for BFKL.
Following the procedure of Ref.~\cite{KMR}, we first neglect the subleading
$k_t \sim \mu$ and $z \sim 1$ effects to obtain and solve a unified BFKL/DGLAP
equation for $h (x, k_t^2)$.  Then, in the last step of the evolution, we take
account of the precise kinematics so that the $k_t \sim \mu$ and $z \sim 1$
domains are treated correctly.

The unified equation for $h_g$, which closely follows that presented in
Ref.~\cite{KMS}, takes the form
\bea   
\label{eq:a7}
h_g (x, \mu^2) & = & h_g^0 (x, \mu^2) \:  + \:  \frac{\alpha_S (\mu^2)}{2 \pi}
\int_0^1 dz \int_{k_0^2}^{\mu^2} \:  \frac{dk_t^{\prime 2}}{k_t^{\prime 2}}
\left [ \Theta (z - x) \:  \bar{P} (z) \:  h_g \left ( \frac{x}{z},
k_t^{\prime 2} \right ) \right .  \nonumber \\ & & \nonumber 
\\ & & - \;
zP_{gg} (z) \:  h_g (x, k_t^{\prime 2}) \:  + \:  \Theta (z - x) \:  P_{gq}
(z) \: \sum h_q \left .  \left (\frac{x}{z}, k_t^{\prime 2} \right ) \:  - \:
P_{gq} (z) \: \sum h_q (x, k_t^{\prime 2}) \right ] \nonumber \\ & & \nonumber \\
& & + \; \frac{\alpha_S (\mu^2)}{2 \pi} \:  2N_C \int_x^1 \frac{dz}{z}
\int_{k^2_0}\frac{dq^2}{q^2} \:  \Theta (k_t^{\prime 2} - k_0^2) \:  \left [
\Theta (\mu^2 - zq^2) \:  \frac{\mu^2}{k_t^{\prime 2}} \:  h_g \left (
\frac{x}{z}, k_t^{\prime 2} \right ) \right .  \nonumber \\ & & \nonumber \\ &
& - \; \left .  \Theta (\mu^2 - q^2) \:  h_g \left (\frac{x}{z}, \mu^2 \right
) \right ],
\eea   
where $\mbox{\boldmath $k$}_t^\prime = \mbox{\boldmath $k$}_t + (1 - z) 
\mbox{\boldmath $q$}$, see Fig.~\ref{fig:vars}.

We have introduced single-scale unintegrated quark auxiliary functions
$h_q(x,\mu^2)$ on the same footing as $h_g$,   
\be  
\label{eq:a8}
h_q (x, \mu^2) \; = \;  \frac{\partial (xq (x, \mu^2))}{\partial \ln
\mu^2},
\ee
and in (\ref{eq:a7}) we sum over all $2n_F$ active flavours $q$
of quarks and antiquarks with
$m_q < \mu$.  The $h_q$ distributions satisfy the equation
\bea   
\label{eq:a9}
h_q (x, \mu^2) & = & h_q^0 (x, \mu^2) \:  + \:  \frac{\alpha_S (\mu^2)}{2 \pi} \:
\int_0^1 \:  dz \:  \int_{k_0^2}^{\mu^2} \:  \frac{dk_t^{\prime 2}}{k_t^{\prime
2}} \:  \left \{ \Theta (z - x) \left [ \:  P_{qg} (z) \:  h_g \left
(\frac{x}{z}, k_t^{\prime 2} \right ) \right .  \right .  \nonumber \\ & &
\nonumber 
\\ & & + \; \left .  \left .  P_{qq} (z) \:  h_q \left ( \frac{x}{z},
k_t^{\prime 2} \right ) \right ] \:  - \:  P_{qq} (z) \:  h_q (x, k_t^{\prime 2})
\right \},
\eea   
which is the usual DGLAP equation for quark evolution written in
terms of $h_q$ of (\ref{eq:a8}).  The last term of (\ref{eq:a7}) is the BFKL
contribution which sums up all the leading $(\alpha_S \ln 1/x)^n$ terms, while the
remaining terms on the right-hand-side describe the conventional DGLAP evolution
of the gluon distribution with respect to scale $\mu^2$.  To avoid double
counting, we have excluded the singular part of the $P_{gg}$ splitting function in
the real emission DGLAP term and used
\be   
\label{eq:a10}   
\bar{P}(z) \; = \; P_{gg} (z) \: - \: \frac{2N_C}{z}.   
\ee 
The $2N_C/z$ term is already included in the BFKL contribution to (\ref{eq:a7}).

The driving terms, $h^0$, which describe the low $k_t^2 < k_0^2$ domain are given
by \cite{KMS}   
\bea   
\label{eq:a11}
h_g^0 (x, \mu^2) & = & \frac{\alpha_S (\mu^2)}{2 \pi} \:  \int_0^1 \:  dz \left \{
\Theta (z - x) \:  \left [ P_{gg}(z) \:  \frac{x}{z} \:  g \left (\frac{x}{z},
k_0^2 \right ) \right .  \right .  \nonumber \\ & & \nonumber \\ & & + \:  \left .
\left .  P_{gq}(z) \: \sum \frac{x}{z} q\: \left (\frac{x}{z}, k_0^2 \right )
\right ] \:  - \:  zP_{gg}(z) \:  xg (x, k_0^2) \:  - \:  P_{gq}(z) \: \sum x
q (x, k_0^2) \right \}, \nonumber \\
\eea
and similarly for $h_q^0$. The integrated input distributions $a (x, k_0^2)$ are
not known and, as usual, must be determined from the data or from some
non-perturbative QCD model.

We emphasize that in the unified BFKL/DGLAP equation we choose the scale $\mu$ to
be $k_t$ for the DGLAP contribution, to be consistent with the BFKL term (which is
independent of $\mu$ at LO).  Recall  that
angular ordering leads to a redefinition of the scale \cite{CCFM,KKMS}
\[   
\Theta (\theta - \theta^\prime) \;\: \Rightarrow \;\: \mu \; > \; z k_t/(1 - z).
\]
If this modified scale were to be adopted then it would be impossible to obtain a
simple one-scale unified evolution equation throughout the whole $x, k_t^2$
domain.  On the other hand, within the LO framework, we may omit the $z$
dependence in the scale, so that the DGLAP part of the evolution becomes ordered
in transverse momenta.  We stress again that it is sufficient to implement the
precise constraints coming from angular ordering (relevant to the $z \sim 0$ and 1
domains for the BFKL and DGLAP terms respectively) at the last step of the
evolution.

An advantage of the single-scale unified BFKL/DGLAP equation is that it is
straightforward to incorporate a major (kinematical) part of the subleading order
$\ln (1/x)$ (BFKL) effects\footnote{The large NLO BFKL corrections, which have
recently been computed \cite{FADIN}, appeared to have put the application of the
BFKL framework into question.  However a major part of the corrections is
kinematic in origin and, when summed to all orders \cite{KMS1} using the theta
function $\Theta (\mu^2 - zq^2)$, brings the BFKL approach back under control, see
also \cite{AND,SALAM}.}  by imposing a consistency condition to ensure that the
virtuality of the exchanged gluon is dominated by its transverse momentum squared
\cite{KMS1}.  This is achieved by the inclusion of the theta function $\Theta
(\mu^2 - zq^2)$ in the real emission contribution shown in the last term of
(\ref{eq:a7}).  Note that other subleading effects arising from using the complete
DGLAP splitting function and running $\alpha_S$ are automatically included in the
unified equation.

We see that the evolution equations for the auxiliary distributions $h_a (x,
k_t^2)$ depend on the single scale $k_t^2$.  The dependence of the unintegrated
distributions $f_a (x, k_t^2, \mu^2)$ on the second scale $\mu$ will enter when we
consider the {\it last step of the evolution}.  It is sufficient to ensure that
the final emitted parton explicitly satisfies the requirements of angular
ordering.  The angular ordering conditions of the previous steps of the evolution
are automatically ensured at LO by virtue of either the strong ordering in $k_t$
(in the DGLAP part) or the strong ordering in $z$ (in the BFKL part).

To ensure angular ordering in the last
step of the evolution, we note that $z$ is limited by (\ref{eq:a12}).  This
condition implies $z < \mu/(\mu + k_t)$, and so we take this as the upper
limit of the $z$ integration in (\ref{eq:a14}) and (\ref{eq:a15}) below.
Thus the number of gluons produced at the last
step (with transverse momentum $k_t$ which initiate a hard subprocess with
factorization scale $\mu$) is\footnote{The low $k_t<k_0$ domain of the integrals
in (\ref{eq:a14}), (\ref{eq:a15}) should be treated as the driving terms $h^0$ in
(\ref{eq:a11}).}
\bea   
\label{eq:a14}  
f_g (x, k_t^2, \mu^2) & = & T_g (k_t, \mu) \:  \frac{\alpha_S (k_t^2)}{2 \pi}
\left \{\int_x^{\mu/(\mu + k_t)} \:  dz \:  \int^{k_t^2} \:  \frac{dk_t^{\prime
2}}{k_t^{\prime 2}} \:  \left [\bar{P}(z) \:  h_g \left (\frac{x}{z}, k_t^{\prime
2} \right ) \right .  \right .  \nonumber \\ & & \nonumber \\ & & + \; \left .
P_{gq}(z) \: \sum h_q \left (\frac{x}{z}, k_t^{\prime 2} \right ) \right ] 
\:  + \:
2N_C \:  \int_x^{\mu/(\mu + k_t)} \:  \frac{dz}{z} \:  \int \:  \frac{d^2 q}{\pi
q^2} \nonumber \\ & & \nonumber \\ & & \left .  \left [ \frac{k_t^2}{k_t^{\prime
2}} \:  h_g \left ( \frac{x}{z}, k_t^{\prime 2} \right ) \:  - \:  \Theta (k_t^2 -
q^2) \:  h_g \left (\frac{x}{z}, k_t^2 \right ) \right ] \right \}, \eea
and the number of a particular quark species $q$, produced at the last
step, is   
\bea   
\label{eq:a15}
f_q (x, k_t^2, \mu^2) & = & T_q (k_t, \mu) \:  \frac{\alpha_S (k_t^2)}{2 \pi} \:
\int_x^{\mu/(\mu + k_t)} \:  dz \:  \int^{k_t^2} \:  \frac{dk_t^{\prime
2}}{k_t^{\prime 2}} \:  \left [  P_{qg}(z) \:  h_g \left (\frac{x}{z},
k_t^{\prime 2} \right ) \right .  \nonumber \\ & & \nonumber \\ & & \hspace{2in} +
\:  \left .  P_{qq}(z) \:  h_q \left ( \frac{x}{z}, k_t^{\prime 2} \right ) \right
].
\eea
This last step of the evolution is shown schematically in Fig.~\ref{fig:laststep}.
The unintegrated distributions $f_a (x, k_t^2, \mu^2)$ represent the number of
partons $\delta a$ emitted in the $\delta \ln k_t^2$ interval from $k_t^2$ to
$k_t^2 + \delta k_t^2$, and include the factors $T_a (k_t, \mu)$ of (\ref{eq:a3}),
which give the probabilities that partons $a = g, q$ with transverse momentum
$k_t$ remain untouched in the DGLAP evolution up to the factorization (probe)
scale $\mu$.  These survival probabilities $T_g$ and $T_q$ resum the virtual DGLAP
contributions occurring in (\ref{eq:a7}) and (\ref{eq:a9}) respectively.

It is important to note that the function $h_g (x, k_t^2)$ already includes the
leading $\ln (1/x)$ virtual corrections, which have the effect of reggeizing the
exchanged gluon; thus $k_t$ is the total momentum transferred via the Regge gluon
trajectory.  Moreover, the distributions $f_a (x, k_t^2, \mu^2)$, evolved in the
final step from the auxiliary functions $h_a (x, k_t^2)$, also incorporate the
virtual DGLAP contributions, via the survival probabilities $T_a (k_t, \mu)$.  The
final expressions, (\ref{eq:a14}) and (\ref{eq:a15}), are thus more symmetric in
that all the LO virtual corrections are included.  That is, from a Feynman diagram
viewpoint, the function $f_a (x, k_t^2, \mu^2)$ corresponds to the propagator of a
$t$-channel parton of transverse momentum $k_t$, initiating a hard subprocess at
scale $\mu$, in which all the LO virtual corrections to the parton distribution
have been taken into account.

\section{The unintegrated gluon distribution}
   
We have described how to obtain the (two-scale) unintegrated parton distributions
$f (x, k_t^2, \mu^2)$ from the solution $h (x, k_t^2)$ of a one-scale equation
which unifies DGLAP and BFKL evolution.  The link is (\ref{eq:a14}),
(\ref{eq:a15}), which represent the last step of the evolution.  Only at this
stage does the scale $\mu$ of the subprocess, initiated by $f (x, k_t^2, \mu^2)$,
enter.

It is informative to compare the unintegrated gluon distribution $f_g (x, k_t^2,
\mu^2)$ with the behaviour of the auxiliary function $h_g (x, k_t^2)$.  Consider
first the pure BFKL limit, in which DGLAP evolution is neglected, that is
$\alpha_S \ln \mu^2 \ll 1$, but $\alpha_S \ln 1/x \gapproxeq 1$.  Then, in
(\ref{eq:a14}), the survival probability $T_g (k_t, \mu) = 1$ and there are no
DGLAP contributions.  Thus we obtain
\be   
\label{eq:a16}
f_g (x, k_t^2, \mu^2) \; = \; h_g (x, k_t^2) \; = \; \frac{\partial (xg (x,
k_t^2))}{\partial \ln k_t^2}.
\ee
This is an expression which is frequently used at low $x$.

When $x$ is sufficiently large, the derivative   
\be   
\label{eq:a17}   
h_a \; = \; \frac{\partial (a (x, k_t^2))}{\partial \ln k_t^2}   
\ee
becomes negative, even for $k_t \lapproxeq \mu$.  The reason is that the negative
virtual DGLAP term exceeds the real emission DGLAP contribution, which is
suppressed by the large lower limit $z > x$ in 
(\ref{eq:a7}), (\ref{eq:a9}), (\ref{eq:a14}) and (\ref{eq:a15}).
After the virtual contributions are resummed into the $T_a$ factors the
unintegrated parton distributions remain positive everywhere.

We already stated in Section~2 that imposing angular ordering on the last step
of ``DGLAP'' evolution caused the distribution to extend smoothly into
the $k_t > \mu$ domain. The gluons also populate this domain due to
BFKL diffusion in $\ln k_t^2$.  This raises the question of how much
enhancement we find in this domain due to the inclusion of the BFKL
contributions. To investigate this point, we 
compute the unintegrated gluon distribution $f_g(x,k_t^2,\mu^2)$
in two different ways.  

\begin{list}{(\alph{bean})}{\usecounter{bean}\setlength{\rightmargin}{\leftmargin}}
\item In the first approach, the function $h_g(x,k^2)$ of Ref.~\cite{KMS}
is used as the auxiliary function to drive the last-step evolution in
(\ref{eq:a14}) and (\ref{eq:a15}).  This incorporates essentially maximal
BFKL effects.\footnote{In comparison with (\ref{eq:a9}), the evolution
of $h_q$ in Ref.~\cite{KMS} contains some resummation of the BFKL-like leading
$(\alpha_S \ln 1/x)^n$ terms.}
\item In the second case, we use unintegrated ``DGLAP'' partons to
drive the two-scale
unintegrated $f_a$ via (\ref{eq:a4}).
This approach is essentially pure DGLAP, but with 
the crucial modification that the cut-off in (\ref{eq:a3}) and
(\ref{eq:a4}) is motivated by
\emph{angular ordering}.  Here we use the MRST99 \cite{MRST99} set of
partons as input.
\end{list}

Ideally, we should refit to the deep inelastic and related
scattering data and perform 
global \emph{unintegrated} parton 
analyses\footnote{In Section 5, we describe the
theoretical calculation of $F_2$, the most important observable
for constraining parton sets, from the two-scale unintegrated $f_g$ and $f_q$.}
in terms of $f_a(x,k_t^2,\mu^2)$, since now we have added an extra last step
to the evolution.
However, such global analyses are beyond the scope of the present paper.

\begin{figure}\centering
\resizebox{\textwidth}{!}{\includegraphics{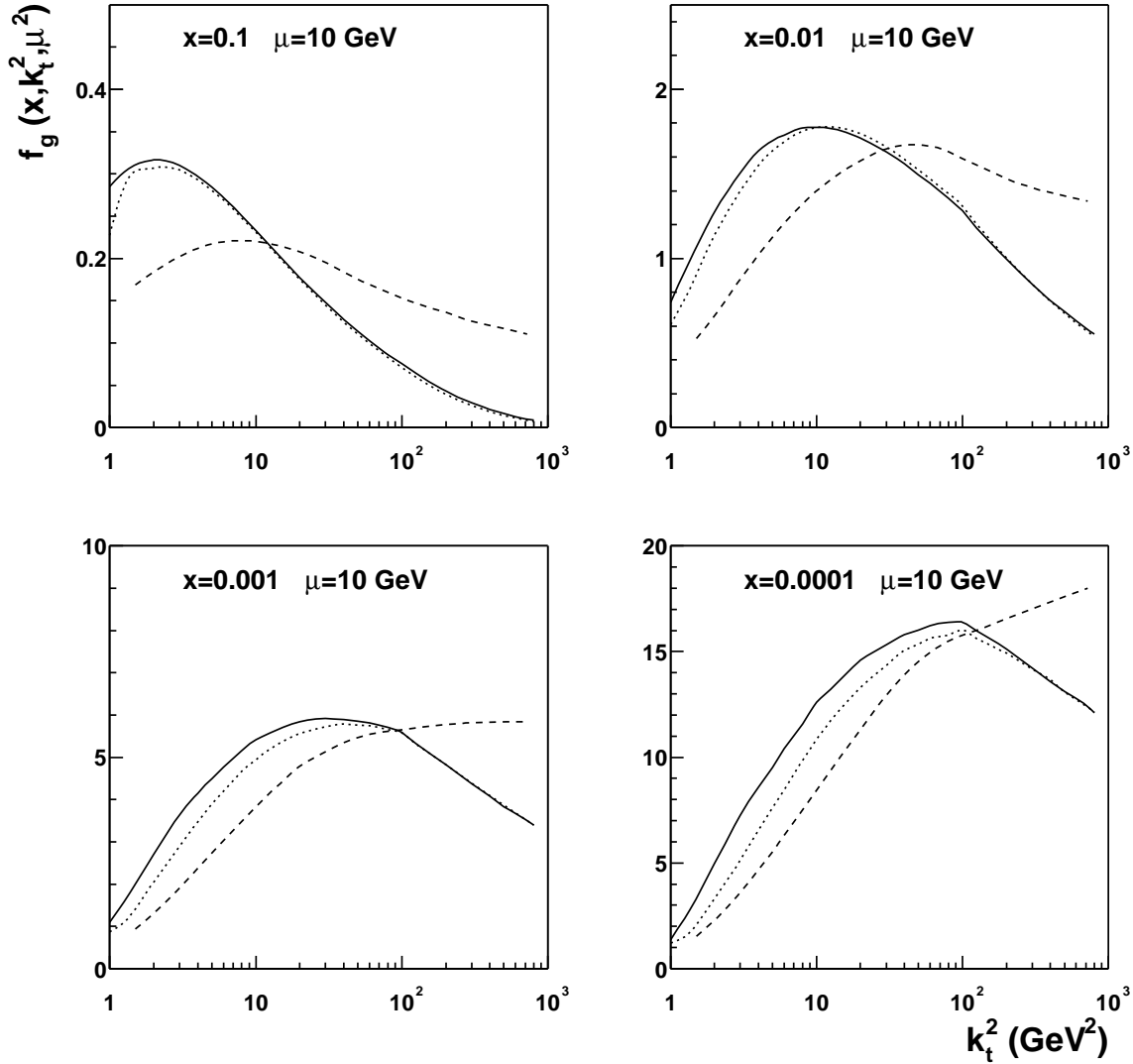}}
\caption{plots of the $k_t$-dependence of the unintegrated gluon
$f_g(x,k_t^2,\mu^2)$ for various values of $x$, at $\mu=10$~GeV.  The solid curves
are our version (a) of 
$f_g$ from (\ref{eq:a14}); for comparison we show with dashed lines the
unintegrated gluon from \cite{KKMS} (as in \cite{KKMS}, the dashed lines 
have been smoothed in the transition region $k_t\sim \mu$).  Also we plot our
``DGLAP'' unintegrated gluon (b) from (\ref{eq:a4}) in dotted lines, which
with the correct angular ordering cut-off is very close to the new $f_g$, 
especially at high $x$.}
\label{fig:unint}
\end{figure}

The unintegrated gluon distribution
$f_g(x,k_t^2,\mu^2)$, obtained by the two alternative procedures (a) and (b),
is shown in Fig.~\ref{fig:unint} for $\mu=10~{\rm GeV}$.  We can compare the 
continuous curves of approach (a) directly with
the previous unintegrated gluon distribution (dashed curves)
calculated in Ref.~\cite{KKMS},
which also used the same auxiliary function $h_g (x, k_t^2)$ of \cite{KMS} as
input.  For $k_t
< \mu$ the new results tend to lie above the previous
 determination \cite{KKMS}, while for
$k_t > \mu$ they are increasingly smaller as $k_t$ increases.  In the present
work, the decrease at large $k_t$ arises from the restriction $z < z_{\rm max} =
\mu/(\mu + k_t)$, whereas the earlier calculation, based on eq.~(23) of
Ref.~\cite{KKMS}, omitted the limit $z < z_{\rm max}$ on the BFKL contribution,
which causes $f_g$ to increase in the large $k_t$ domain at small values of $x$.

Finally we compare the DGLAP-like unintegrated gluon, obtained in approach (b),
with that of approach (a) which embodied BFKL evolution.
That is we compare the
dotted with the continuous curves of Fig.~\ref{fig:unint}.
It is interesting to note that, with the more precise angular-ordered
(CCFM \cite{CCFM})
cut-off from (\ref{eq:a12}), $\Delta = k_t/(\mu + k_t)$, the unintegrated gluon
distribution (b) generated from (\ref{eq:a4}) using the pure DGLAP MRST99
\cite{MRST99} partons turns out to be very similar to the result (a) obtained using
the auxiliary function $h_g$.   
This coincidence can be
explained by the facts (i) that both the DGLAP partons and the analysis of
Ref.~\cite{KMS} (which yielded $h_g$) fit the deep inelastic data well, and (ii)
that both these input are used with the same angular-ordered constraint,
(\ref{eq:a12}).  We draw the conclusion that the role of angular ordering in the
last step of evolution is particularly important, even more so than BFKL
effects in the HERA domain.

\section{\boldmath{$F_2$} calculated from the unintegrated partons}

To check the reliability of our unintegrated parton distributions, and to
demonstrate how to use these distributions in calculations of observables, we
compute the deep inelastic structure function $F_2$.  We wish to treat the
unintegrated gluons and unintegrated quarks on an equal footing as input to the
subprocess cross sections, so we explicitly separate gluon and (direct) quark
contributions to $F_2$.

\begin{figure}\centering
\scalebox{0.7}{\resizebox{\textwidth}{!}{\includegraphics{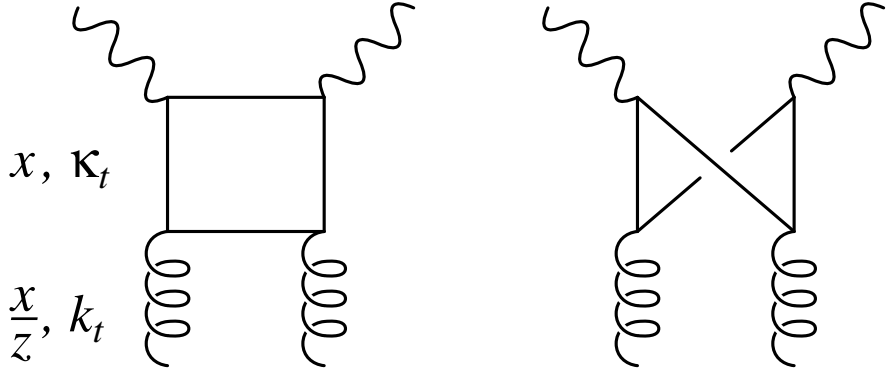}}}
\caption{The quark box, and crossed-box, diagrams which mediate the
contribution of the unintegrated gluon distribution $f_g(x/z,k_t^2,\mu^2)$
to $F_2$.}
\label{fig:boxandcrossedbox}
\end{figure}

The gluon contributes to $F_2$ via the quark box and crossed-box diagrams of
Fig.~\ref{fig:boxandcrossedbox}.  These generate, via the $g \rightarrow q\bar{q}$
splitting, a sea quark contribution $S_q$ to $F_2$ of the form \cite{AKMS,KMS}
\be
\label{eq:a18}
F_2^{g \rightarrow q\bar{q}} \:  (x, Q^2) \; = \; \sum_q \:  e_q^2 \:  S_q (x,
Q^2),
\ee
with
\bea
\label{eq:a19}
& & S_q (x, Q^2) \; = \; \frac{Q^2}{4 \pi^2} \:  \int \:  \frac{dk_t^2}{k_t^4} \:
\int_0^1 \:  d\beta \:  \int \:  d^2 \kappa_t \:  \alpha_S (\mu^2) \:  f_g \left (
\frac{x}{z}, k_t^2, \mu^2 \right ) \:  \Theta \left (1 - \frac{x}{z} \right )
\nonumber \\ & & \\ & & \quad \left \{ \left [ \beta^2 + (1 - \beta)^2 \right ] \:
\left ( \frac{\mbox{\boldmath $\kappa$}_t}{D_1} \:  - \:  \frac{\mbox{\boldmath
$\kappa$}_t - \mbox{\boldmath $k$}_t}{D_2} \right )^2 \:  + \:  \left [m_q^2 +
4Q^2 \beta^2 (1 - \beta)^2 \right ] \:  \left ( \frac{1}{D_1} - \frac{1}{D_2}
\right )^2 \right \}.  \nonumber
\eea
The denominator factors are
\bea
\label{eq:a20}
D_1 & = & \kappa_t^2 \:  + \:  \beta (1 - \beta) \:  Q^2 \:  + \:  m_q^2 \nonumber
\\ & & \\ D_2 & = & (\mbox{\boldmath $\kappa$}_t - \mbox{\boldmath $k$}_t)^2 \:  +
\:  \beta (1 - \beta) \:  Q^2 \:  + \:  m_q^2.  \nonumber
\eea
We may exploit the symmetry of the integrand in (\ref{eq:a19}) under
$\mbox{\boldmath $\kappa$}_t \rightarrow \mbox{\boldmath $\kappa$}_t -
\mbox{\boldmath $k$}_t$ and $\beta \rightarrow 1 - \beta$ to rewrite $\{ \ldots
\}$ as
\be
\label{eq:a21}
2 \:  \left \{ \left [ \beta^2 + (1 - \beta)^2 \right ] \:  \left (
\frac{\kappa_t^2}{D_1^2} \:  - \:  \frac{\kappa_t^2 - \mbox{\boldmath $\kappa$}_t
\cdot \mbox{\boldmath $k$}_t}{D_1 D_2} \right ) \:  + \:  \left [m_q^2 \:  + \:
4Q^2 \beta^2 (1 - \beta)^2 \right ] \:  \left ( \frac{1}{D_1^2} \:  - \:
\frac{1}{D_1 D_2} \right ) \right \}.
\ee
The summation in (\ref{eq:a18}) is over massless $u, d, s$ quarks and a $c$ quark
of mass $m_c = 1.4~{\rm GeV}$; there is no need to sum $\bar{u}$, $\bar{d}$,
$\bar{s}$, $\bar{c}$ in addition, because as (\ref{eq:a19}) is written,
$S_u$, say, is the contribution of a gluon via a $u$ quark box of
any momentum.  The variable $\beta$ is the light-cone fraction of
the photon momentum carried by the internal quark.  The variable $z$ is the ratio
of Bjorken $x$ and the fraction of the proton momentum carried by the gluon.  It
is specified by the relation
\bea
\label{eq:a22}
\frac{1}{z} & = & 1 \:  + \:  \frac{(\mbox{\boldmath $\kappa$}_t - (1 - \beta)
\mbox{\boldmath $k$}_t)^2 + m_q^2}{\beta (1 - \beta) Q^2} \:  + \:
\frac{k_t^2}{Q^2} \nonumber \\ & & \\ & = & 1 \:  + \:  \frac{\kappa_t^2 +
m_q^2}{(1 - \beta) Q^2} \:  + \:  \frac{k_t^2 + \kappa_t^2 - 2 \mbox{\boldmath
$\kappa$}_t \cdot \mbox{\boldmath $k$}_t + m_q^2}{\beta Q^2}, \nonumber
\eea
which is obtained by requiring the outgoing quarks to be on-shell.  Following
Ref.~\cite{KMS}, we choose the scale $\mu$ which controls the unintegrated gluon
distribution and the QCD coupling $\alpha_S$ to be
\be
\label{eq:a23}
\mu^2 \; = \; k_t^2 \: + \: \kappa_t^2 \: + \: m_q^2.
\ee

Care is needed in separating this calculation into perturbative and
non-perturbative regions.  We impose a cut-off $k_t > k_0$ for
a legitimate perturbative calculation of (\ref{eq:a19}).  The smallest
cut-off we can choose is the minimum (that is, initial) scale of the
function $h_g(x,k_t^2)$ from which our two-scale distributions derive.
Thus $k_0$ is of order 1 GeV. For
the contribution from the region of $k_t < k_0$ we approximate
\be
\label{eq:a24}
\int_0^{k_0^2} \:  \frac{dk_t^2}{k_t^2} \:  f_g (x, k_t^2, \mu^2) \:  \left [
\frac{\rm remainder}{k_t^2} \right ] \; \simeq \; xg (x, k_0^2) \:  T_g (k_0, \mu)
\:  \Bigg[ \quad \Bigg]_{k_t = a},
\ee
where $a$ can be taken to be any value representative of the interval $(0, k_0)$.
The dependence on the choice of $a$ is numerically unimportant.

Now we have to add the direct quark contributions to $F_2$, which come from the
unintegrated quark distributions $f_q (x, k_t^2, \mu^2)$.  If a quark, initially
with $x/z$ and perturbative transverse momentum $k'_t>k_0$, splits to a 
radiated gluon and a `final'\footnote{By `final' we mean the struck quark
 just after the $P_{qq}$ splitting.} quark with Bjorken $x$ and
 transverse momentum
$\kappa_t$, then this final quark can couple to the photon and contribute to
$F_2$ as:
\begin{eqnarray}
\nonumber\lefteqn{F_2^{q ({\rm pert})} \:  (x, Q^2) \; = \; \sum_q \:  e_q^2 \:  \int_{k_0^2}^{Q^2}
\:  \frac{d \kappa_t^2}{\kappa_t^2} \:  \frac{\alpha_S (\kappa_t^2)}{2 \pi} \:
\int_{k_0^2}^{\kappa_t^2} \:  \frac{dk_t^2}{k_t^2} \:  \int_x^{Q/(Q + k_t)} \:  dz}\\
&&\nonumber\\
&&\hspace{7cm} \left[ f_q \left ( \frac{x}{z}, k_t^2, Q^2 \right ) 
              \: + f_{\bar{q}} \left ( \frac{x}{z}, k_t^2, Q^2 \right ) \right]
              \:  P_{qq} (z).\label{eq:a25}
\end{eqnarray}
where here we have written the antiquark contribution explicitly.
As in (\ref{eq:a15}), the upper limit of the $z$ integration reflects the
angular-ordered constraint of (\ref{eq:a12}) during the quark evolution.

Again we need to account for the non-perturbative domain $k_t<k_0$.  The initial
(integrated) quark distribution $xq(x,k_0^2)$ drives our final contribution.
Physically the only remaining diagrams that we have not included are those
in which a quark (or antiquark)
from this initial distribution does not experience real splitting in
the perturbative domain, but interacts
unchanged with the photon at scale $Q$.  Hence
we write a Sudakov-like factor $T_q(k_0,Q)$ to represent the probability
of evolution from $k_0$ to $Q$ without radiation.
\be
\label{eq:a26}
F_2^{q ({\rm non-pert})} \:  (x, Q^2) \; = \; \sum_q \:  e_q^2 \left (xq (x,
k_0^2) \:  + \:  x\bar{q} (x, k_0^2) \right ) \:  T_q (k_0, Q).
\ee
To avoid double counting, it
is important to put a lower limit on $\kappa_t$ in both
(\ref{eq:a19}) and (\ref{eq:a25}), by
enforcing  $\Theta (\kappa_t^2 - k_0^2)$ in the integrations.  Without this
lower limit on the final transverse momentum $\kappa_t$, equations
(\ref{eq:a19}) and (\ref{eq:a25}) would partially 
include low transverse momentum $\kappa_t$ quark contributions
which are best incorporated in (\ref{eq:a26}), whether they originate
from partons with $k_t>k_0$ or not.

\begin{figure}\centering
\resizebox{\textwidth}{!}{\includegraphics{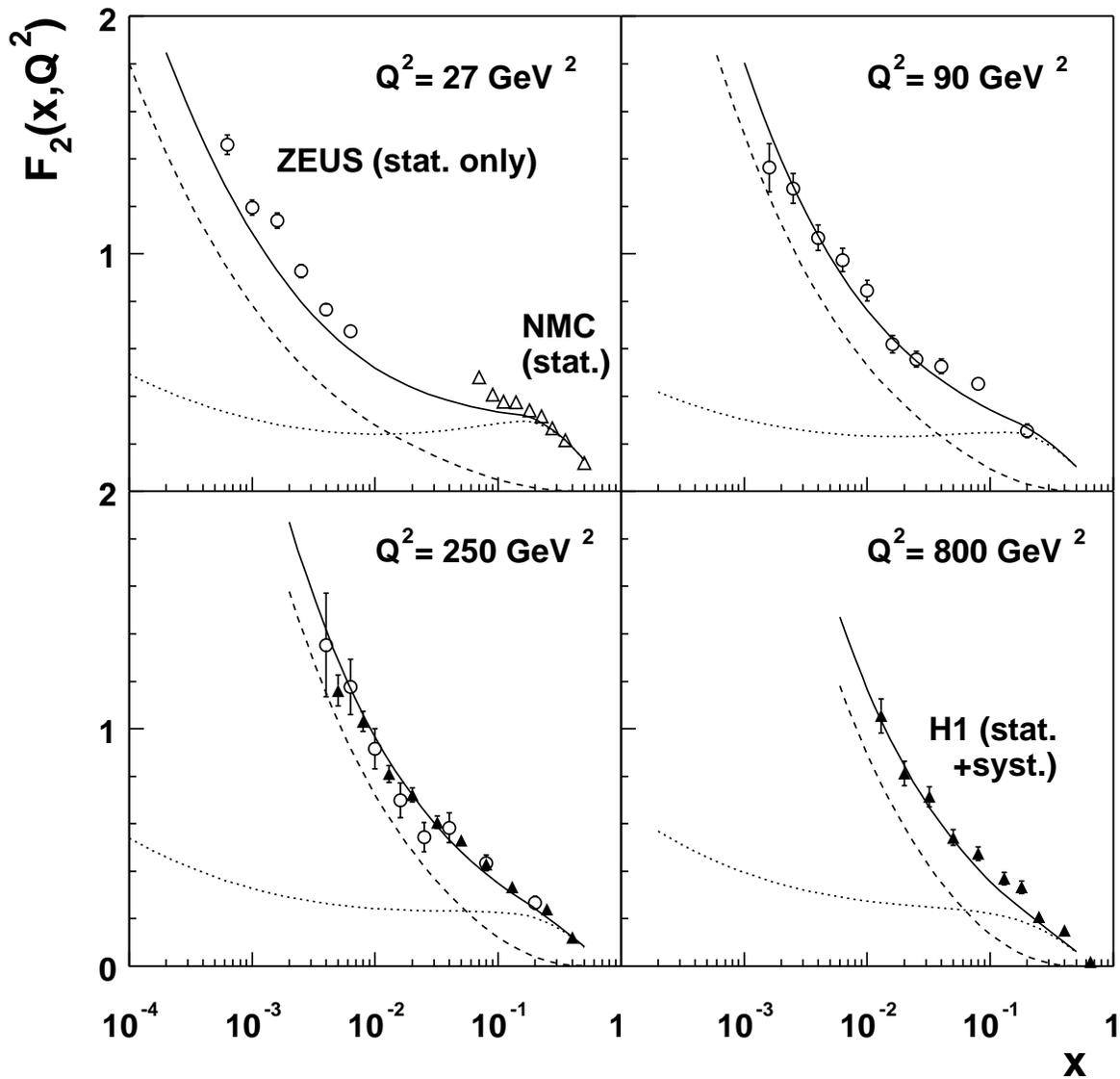}}
\caption{This is not a fit but the results of using our ``DGLAP'' 
unintegrated partons (b) to calculate $F_2$; the gluon-originated
contributions are shown as dashed lines and the quark-originated parts
are shown as dotted lines.  Recent data are plotted \cite{DATA},
and compare well with the sum of the gluon and quark contributions
(solid curves), especially at high $Q^2$.}
\label{fig:fourq2}
\end{figure}

The structure function $F_2 (x, Q^2)$ is given by the sum of the gluon-initiated
contribution (\ref{eq:a18}), and the quark terms, (\ref{eq:a25}) and
(\ref{eq:a26}).  In Fig.~\ref{fig:fourq2} sample results\footnote{We show
in Fig.~\ref{fig:fourq2} results obtained using the unintegrated distributions
evaluated from (\ref{eq:a4}) with the MRST99 partons \cite{MRST99}, that is,
version (b) of $f_g$ and $f_q$ as discussed in Section~4.}, shown by the
continuous curves, are compared with deep-inelastic structure function data.  The
gluon and quark components are shown by the dashed and dotted curves respectively.
As expected the dominant contribution at small $x$ comes from the unintegrated
gluon via the quark box and crossed-box contributions, whereas at large $x$ the
quark terms dominate.

We emphasize that, in the present work, the curves for $F_2$ are not the result of
a fit to the structure function data.  Rather they have been obtained by using
single-scale functions, originally fitted more directly to $F_2$ data,
as plausible input to our `last-step' evolution procedure, which generates
two-scale unintegrated distributions $f_a (x, k_t^2, \mu^2)$.  Then we use the 
unintegrated distributions to compute $F_2$ via
(\ref{eq:a18}), (\ref{eq:a25}) and (\ref{eq:a26}).  For the inclusive observable
$F_2$, we would expect that the insertion of this extra evolution step would not
appreciably disturb the description, since it essentially redistributes the
distributions in $k_t$ space.  We see from Fig.~\ref{fig:fourq2} that indeed this
is the case.

\section{Relation of \boldmath{$f$} to integrated partons}

It is important to scrutinise the relationship between the new
unintegrated partons $f_a(x,k_t^2,\mu^2)$ and the conventional
integrated parton distributions $a(x,\mu^2)$, as obtained in
global analyses such as
\cite{MRST99}. First we emphasize that we may use either the
integrated distributions or the unintegrated distributions to describe
both inclusive (like $F_2$) and exclusive processes.  The framework
based on the unintegrated distributions is a bit more complicated.
However it accounts for the precise kinematics of the process and an
important part of the virtual loop corrections, via the survival factor
$T$, even at LO.  On the other hand, if we work with integrated
partons we have to include NLO (and sometimes NNLO) contributions
to account for these effects.  These differences appear to cause a
discrepancy between
the integrated and unintegrated approaches.  As we explain below,
this is to be expected since it arises from
simplifications of the LO formalism due to the neglect of terms
which are moved into the NLO contribution.

An important equation, sometimes cited as the defining 
property of unintegrated partons \cite{KKMS}, is
\begin{equation}\label{eq:intofunint}
a(x,\mu^2)=\int^{\mu^2} \frac{dk_t^2}{k_t^2} \, f_a\left(x,k_t^2,\mu^2\right),
\end{equation}
where $a$ represents $xg$ or $xq$.  This is in fact the first equation of
Ref.~\cite{KKMS}.  In the BFKL limit, the $\mu$ dependence of $f$ vanishes and
we have $f_g(x,k_t^2,\mu^2)\rightarrow h_g(x,k_t^2)$ as in
(\ref{eq:a16}).  In this case, (\ref{eq:intofunint}) is clearly satisfied.
However, in general the situation is complicated by the two separate
momentum scales $k_t$ and $\mu$.  The unintegrated partons $f_g$ of 
Ref.~\cite{KKMS} were explicitly constructed to have the property 
(\ref{eq:intofunint}), in the sense that the integral of $f_g$ over the
transverse momentum up to the scale $\mu$ would be the same as the
integral of the input auxiliary function $h_g(x,k_t^2)$ 
up to the same scale.  In contrast, numerical integration
over $k_t$ of the new unintegrated partons $f_g$ and $f_q$ presented
in this paper (both versions (a) and (b) of Section~4)
shows that (\ref{eq:intofunint})
is only approximately true.\footnote{Note that we cannot compute
(\ref{eq:intofunint}) as it is written, because we can only define
the unintegrated function in the regime of perturbative $k_t > k_0$.
The comparison that is made is between the integral from $k_0^2$ to $\mu^2$
and the quantity $a(x,\mu^2)-a(x,k_0^2)$.}  We typically find a
discrepancy of order 25\% between the right-hand side of
(\ref{eq:intofunint}) and the single-scale distribution that has been used to
generate $f_a$.

In order to eliminate the discrepancy we may adjust the upper limit 
$\mu^2$ of the integral in (\ref{eq:intofunint}) to $c^2\mu^2$.  The
introduction of $c$ is equivalent to a NLO correction for the integrated
partons.  Typically in the ``DGLAP'' case (approach (b) of
Section~4) we require $c=0.6-0.8$ to reproduce\footnote{This
identification is only approximate since the last step of the
evolution is not included in the comparison.} the original MRST
integrated gluon in the domain $\mu=5-10\ {\rm GeV}$ and $x\lapproxeq
0.01$ (and $c\approx 0.4$ for approach (a), which embodies BFKL effects).
A value $c<1$ compensates for the over-large virtual loop correction
included in the integrated DGLAP partons on account of the absence
of the cut-off $\Delta$ in the integral over $z'$.  At smaller
values of $x$ the corrections due to the cut-off are mainly cancelled
between the virtual and real DGLAP contributions.  As $x$ increases,
the virtual contribution (the second term on the right-hand-side of
(\ref{eq:a1})) increasingly dominates and we have to choose smaller
values of $c$.  Eventually in the domain $x>0.1$, $\mu\sim 10\ {\rm GeV}$
the main contribution comes from the input and to compensate the
absence of the $z$ cut-off in the conventional form of the DGLAP
equation we must change the input itself.

To summarize, the discrepancy between the integral (\ref{eq:intofunint})
of the unintegrated parton function and the 
original integrated distribution is 
not a cause for concern.  Conceptually,
there are two different roles for single-scale distributions in the
description of data for inclusive observables (where partonic transverse
momentum is integrated out).  The first role is the traditional one,
in the framework of collinear factorization, whereby integrated parton
distribution functions are fitted directly to the data.  The second role
is demonstrated in this paper (following \cite{KMR} and \cite{KKMS}), where
we use single-scale functions as \emph{input} to the last-step procedure, see
for example (\ref{eq:a4}).  We have been forced to introduce
a new formalism for
calculating $F_2(x,Q^2)$, in which unintegrated parton functions are
understood to be the fundamental objects; we have emphasized the need
to perform a new global fit to data in terms of the new functions
$f_a$.  After this, we do not expect
the input single-scale function
$a$ on the left-hand side of (\ref{eq:intofunint}) to equal the integral
of $f_a$ up to $\mu^2$, since $a$ itself is not fitted directly to the data, but rather
is used as input for the last step of the evolution, which embodies
a crucial angular-ordering constraint unique to this last step.
Thus our single-scale or `auxiliary'
function is not a traditional parton distribution function, but 
simply an intermediate function.

\section{Conclusions}

Parton distributions, $f_a (x, k_t^2, \mu^2)$ unintegrated over the parton $k_t$
are the basic quantities for describing processes initiated by hadrons.
An essential ingredient in this description is the existence of the $k_t$
factorization theorem \cite{KTFAC}.  The unintegrated distributions depend on two
hard scales --- the transverse momentum $k_t$ and the factorization scale $\mu$.
The scale $\mu$ drives the angular ordering during the evolution which arises from
the coherence of the gluon emissions.

Here we develop a new formalism to determine the unintegrated parton
distributions, $f_a (x, k_t^2, \mu^2)$, which embodies both the leading $\ln Q^2$
(DGLAP) and $\ln 1/x$ (BFKL) effects, as well as including a major part of the
sub-leading $\ln 1/x$ contributions.  An important observation is that, at
leading order, the {\it two-scale} functions $f_a (x, k_t^2, \mu^2)$ may be
calculated from auxiliary functions $h_a (x, k_t^2)$ which satisfy {\it
single-scale} evolution equations, since the angular ordering restrictions,
controlled by $\mu$, become important only at the last step of the evolution.  The
equation for the auxiliary function $h_g (x, k_t^2)$ was formulated, and the
distributions fitted to the data, in Ref.~\cite{KMS}.  These single-scale
equations were also devised to include all the leading $\alpha_S \ln Q^2$ and
$\alpha_S \ln 1/x$ contributions, and a major part of the sub-leading $\ln 1/x$
effects.

In other words, the `unified' evolution equations for $h$ must be supplemented by
a final evolution step in which the $\mu$ dependence of the unintegrated
parton distributions enters via the angular-ordering constraint.  The
situation is summarised diagrammatically in Fig.~\ref{fig:laststep}.
  The procedure offers a
considerable simplification in the determination of physically realistic
unintegrated parton distributions $f_a (x, k_t^2, \mu^2)$ over the full $x, \mu^2$
perturbative domain, including the true kinematics even at leading order.  As
expected, the gluon and sea quark distributions extend into the $k_t > \mu$ region
more and more as $x$ decreases.  We have compared the new unintegrated
distributions with those given by previous prescriptions \cite{KMR,KKMS}.  As
compared to Ref.~\cite{KKMS}, the new formalism gives a consistent treatment of
angular ordering, which leads to the imposition of the
$z$ integration limit $z < \mu/(\mu + k_t)$.  As a consequence the distributions
$f_a (x, k_t^2, \mu^2)$ decrease faster than those of \cite{KKMS} for large $k_t$,
particularly at small $x$.  An interesting result is that the unintegrated
distributions obtained via $h_a (x, k_t^2)$ of \cite{KMS} are not very different
from those obtained via (\ref{eq:a4}) using conventional DGLAP partons --- compare
the continuous and dotted curves in Fig.~\ref{fig:unint}.  It thus appears that
the imposition of the angular-ordering constraint is more important than the
BFKL effects.  This observation has the practical
consequence that reasonably accurate predictions for observables can be made using
the much simpler, though less complete, prescription of (\ref{eq:a4}).

Finally, we used the new unintegrated distributions to calculate the
deep-inelastic structure function $F_2$.  We also showed the gluon-initiated and
quark contributions separately, which, as expected, dominate at small $x$ and
large $x$ respectively.  Recall, from \cite{KMS}, that the rise of the gluon at
small $x$ is driven by perturbative QCD, which was assumed to have a
non-perturbative input which is `flat' in $x$.  We emphasize that we have not
fitted to the deep inelastic data.  Nevertheless, Fig.~\ref{fig:fourq2}
 shows that the existing
distributions give an adequate description, and therefore they may be used to
evaluate other hard processes, such as $b\bar{b}$ and large $q_t$ prompt photon
production in high energy $p\bar{p}$ (or $pp$) collisions.  It is important to use
unintegrated distributions for such exclusive reactions.

\section*{Acknowledgements}

We thank Jan Kwiecinski and Anna Stasto for many valuable discussions during the
course of this research.  The work was supported by the UK Particle Physics and
Astronomy Research Council (PPARC), and also the Russian Fund for Fundamental
Research (98-02-17629).  This work was also supported by the EU Framework TMR
programme, contract FMRX-CT98- 0194 (DG 12-MIHT).

\end{document}